\documentclass[12pt]{article}
\usepackage{graphicx}
\usepackage{makeidx,amsfonts,graphicx,amsthm,amsmath}
\newtheorem{theorem}{Theorem}{\bfseries}{\rmfamily}
\newtheorem{lemma}[theorem]{Lemma}{\bfseries}{\rmfamily}
\newtheorem{proposition}[theorem]{Proposition}{\bfseries}{\rmfamily}
\newtheorem{corollary}[theorem]{Corollary}{\bfseries}{\rmfamily}

\theoremstyle{definition}
\newtheorem*{definition}{{Definition}}{\bfseries}{\rmfamily}
\newtheorem{conjecture}[theorem]{Conjecture}
\newtheorem{algorithm}[theorem]{Algorithm}{\bfseries}{\rmfamily}
\newtheorem{algorithm*}{Algorithm}{\bfseries}{\rmfamily}

\theoremstyle{remark}
\newtheorem*{remark}{{Remark}}{\bfseries}{\rmfamily}
\newtheorem*{example}{{Example}}{\bfseries}{\rmfamily}

\newcommand{\bt}{\begin{tabular}}
\newcommand{\et}{\end{tabular}}
\newcommand{\ba}{\begin{array}}
\newcommand{\ea}{\end{array}}
\newcommand{\bq}{\begin{eqnarray}}
\newcommand{\eq}{\end{eqnarray}}
\newcommand{\bqa}{\begin{eqnarray*}}
\newcommand{\eqa}{\end{eqnarray*}}
\newcommand{\ds}{\displaystyle}

\def\zz{{\mathbb Z}}
\def\nn{{\mathbb N}}
\def\ff{{\mathbb F}}
\def\fqo{\ff_q^\infty}
\def\fqw{\left(\ff_q^M\right)^\infty}
\newcommand{\bthm}{\begin{theorem}}
\newcommand{\ethm}{\end{theorem}}
\newcommand{\bpro}{\begin{proposition}}
\newcommand{\epro}{\end{proposition}}
\newcommand{\blem}{\begin{lemma}}
\newcommand{\elem}{\end{lemma}}
\newcommand{\bp}{\begin{proof}}
\newcommand{\ep}{\end{proof}}
\newcommand{\bcon}{\begin{conjecture}}
\newcommand{\econ}{\end{conjecture}}
\newcommand{\bcor}{\begin{corollary}}
\newcommand{\ecor}{\end{corollary}}
\newcommand{\bde}{\begin{definition}}
\newcommand{\ede}{\end{definition}}
\newcommand{\bexa}{\begin{example}}
\newcommand{\eexa}{\end{example}}

\begin{document}
 
\pagestyle{plain}
{\Large
{\bf \noindent
Towards a General Theory of\\ 
Simultaneous Diophantine Approximation of\\
Formal Power Series:\\
Linear Complexity of Multisequences\\
}

\noindent
Michael Vielhaber\footnote{
Supported by Project FONDECYT 2004, No. 1040975 of CONICYT, Chile}
and 
M\'onica del Pilar Canales Chac\'on\footnotemark[\value{footnote}]\\
}

\noindent
Instituto de Matem\'aticas, Universidad Austral de Chile,\\
Casilla 567, Valdivia,  Chile\ \
{{\rm\{}\tt vielhaber,monicadelpilar{\rm\}}@gmail.com}

\begin{abstract}

We model the development of the linear complexity of multisequences by
a stochastic infinite state machine, the Battery--Discharge--Model,
BDM. The states $s\in S$ of the BDM have asymptotic 
probabilities or mass 
$\mu_\infty(s)={\cal P}(q,M)^{-1}\cdot q^{-K(s)}$, 
where $K(s)\in\nn_0$ is 
the {\it class} of the state $s$, and 
${\cal P}(q,M)=\sum_{K\in\nn_0}P_M(K)q^{-K}=\prod_{i=1}^M   q^i/(q^i-1)$ 
is the generating function of the number of partitions into at most
$M$ parts. We have (for each timestep modulo $M+1$) 
just $P_M(K)$ states of class $K$.

We obtain a closed formula for the asymptotic probability for the
{\it linear complexity deviation} $d(n) := L(n)-\lceil n\cdot
M/(M+1)\rceil$ with 
\[\gamma(d)=\Theta\left(q^{-|d|(M+1)}\right),\forall M\in \nn, \forall
d\in\zz.\] 
The precise formula is given in the text.
It has been verified numerically for $M=1,\dots,8$, and is  
conjectured to hold for all $M\in \nn$.

From the asymptotic growth (proven for all $M\in\nn$), 
we infer the Law of the Logarithm for
the linear complexity deviation,
\[
-\liminf_{n\to\infty}\frac{d_a(n)}{\log n} 
= \frac{1}{(M+1)\log q}
=\limsup_{n\to\infty}\frac{d_a(n)}{\log n}, 
\]
which immediately yields 
$\ds\frac{L_a(n)}{n}\to\frac{M}{M+1}$ with measure one, 
$\forall M\in\nolinebreak\nn,$ 
a result recently shown already by Niederreiter and Wang.

{\bf Keywords:} Linear complexity, linear complexity deviation,
multi\-sequen\-ce, Battery Discharge Model, isometry.

\end{abstract}

\newpage
\subsection*{1. Linear Complexity of Multisequences}

The {\it linear complexity} of a finite string $a\in\ff_q^n$, $L_a(n)$, is the
least length of an LFSR (Linear Feedback Shift Register), which
produces $a_1,\dots,a_n$ starting with an initial content
$a_1,\dots,a_{L_a(n)}$. If all symbols are zero, we set
$L_{(0,\dots,0)}(n) = 0$. Also, we put $L_a(0)=0$ for all $a$.

An alternative and equivalent definition defines $L_a(n)$ as the
length of the shortest
recurrence within the $a_i$, {\it i.e.}

\[L_a(n) := \min_{1\leq l\leq n}\left(
\exists\alpha_1,\dots,\alpha_{l-1}\in\ff_q,
\forall 1\leq k\leq n-l\colon 
a_{k+l}=\sum_{i=1}^{l-1} \alpha_i\cdot a_{k+i}\right).\]

Given an infinite sequence $a\in\fqo$, we define $L_a(n)$ as before,
taking into account only the finite prefix $a_1,\dots,a_n$. The
sequence $\left(L_a(n)\right)_{n\in\nn_0}$ is called the {\it linear complexity
profile} of $a$.
The diophantine approximation of the generating function
$G(a) := \sum_{n=1}^\infty a_nx^{-n}\in \ff_q[[x^{-1}]]$ by a
polynomial function with precision at least $k$ that is
\[G(a) = \frac{u(x)}{v(x)} + o(x^{-k}),\]
requires a polynomial $v(x)$ of degree at least $L_a(k)$, and this
length is also sufficient, since $v(x)$ may be chosen as  the feedback 
polynomial of the LFSR producing $a_1,\dots,a_k$.

Turning to multisequences $(a_{n,m})_{n\in\nn, 1\leq m\leq M} \in
\fqw$, we ask for {\it simultaneously} approximating all $M$ formal
power series 
\[\ff_q[[x^{-1}]]\ni G_m(a) := \sum_{n=1}^\infty a_{n,m}x^{-n} 
=\frac{u_m(x)}{v(x)} + o(x^{-k}),1\leq m\leq M\]
with the {\it same} denominator polynomial $v(x)$, equivalently we
search a single LFSR which produces all $M$ sequences with  suitable
initial contents.
The linear complexity profile of $a$  now is defined by
$(L_a(n,m))_{n\in\nn_0,1\leq m\leq M}$  (symbol by symbol), and we set 
$L_a(n) := L_a(n,M)$, when considering only complete
columns of all $M$ sequences at the same place $n$, with profile
$(L_a(n))_{n\in\nn_0}.$ 

\noindent{\bf 
The goal of this paper is to characterize the behaviour of $L$  
as a probability distribution over {\it all}
multisequences from $\fqw$.
}

\subsection*{2. Continued Fraction Expansion:\\
Diophantine Approximation  of Multisequences}

The task of determining the linear complexity profile of {\it one}
multi\-se\-quen\-ce from $\fqw$ has been resolved by Dai and
Feng \cite{Dai}. Their mSCFA (multi--Strict Continued Fraction
Algorithm) computes  a sequence 
\[\left(\left(\frac{u_k^{(n,m)}}{v^{(n,m)}}\right)_{k=1}^M\right)_{(n,m)
\in \{1,\dots,M\} \times\nn_0} \]
of best simultaneous approximations to $(G_k)$. The order of timesteps is
$(n,m) = (0,M),(1,1),(1,2),\dots,(1,M),(2,1),(2,2),\dots$ with 
$$G_m(a) = \sum_{t\in\nn}a_{m,t}\cdot x^{-t} =
\frac{u_m^{(n,m)}(x)}{v^{(n,m)}(x)}+o(x^{-n}),
n\in\nn_0, \forall\ 1\leq m\leq M.$$ 

We will denote the degree of $v^{(n,m)}(x)$ by $\deg(n,m)\in\nn_0$, 
thus the linear complexity profile is
 $(\deg(n,M))_{n\in\nn_0}= (L_{(G_m,1\leq m\leq M)}(n))_{n\in\nn_0}.$

The mSCFA  uses $M$ auxiliary degrees $w_1,\dots,w_M\in\nn_0$.
The update of these values (and $\deg$) depends on a so--called 
``discrepancy'' $\delta(n,m)\in\ff_q$.
$\delta(n,m)$ is zero if the current approximation predicts 
correctly the value $a_{n,m}$, and $\delta(n,m)$ is nonzero otherwise.

Furthermore, the polynomials $u_m(x)$ and $v(x)$  are updated, 
crucial for the mSCFA, but of no importance for our concern, and we
omit the respective part of the mSCFA in the program listing:

\begin{algorithm}
{\tt mSCFA}\\
$\deg:=0; w_m:=0,1\leq m\leq M$\\
{\tt FOR} $n := 1,2,\dots$\\
\hspace*{6 mm}    {\tt FOR} $m:=1,\dots,M$\\
\hspace*{12 mm}        {\tt compute} $\delta(n,m)$\ \ //discrepancy\\
\hspace*{12 mm}        {\tt IF} $\delta(n,m) = 0$: \{\} // do nothing, 
\cite[Thm.~2, Case 2a]{Dai}\\
\hspace*{12 mm}        {\tt IF} $\delta(n,m) \neq 0$ {\tt AND}
 $n-\deg-w_m\leq 0: \{\}$ 
 // \cite[Thm.~2, Case 2c]{Dai}\\
\hspace*{12 mm}        {\tt IF} $\delta(n,m) \neq 0$ {\tt AND}
 $n-\deg-w_m > 0$:  
 // \cite[Thm.~2, Case 2b]{Dai}\\
\hspace*{18 mm}          $\mbox{deg\_copy}:=\deg$\\
\hspace*{18 mm}          $\deg := n -w_m$\\
\hspace*{18 mm}          $w_m := n - \mbox{deg\_copy}$\\
\hspace*{6 mm}     {\tt ENDFOR}\\
{\tt ENDFOR}
\end{algorithm}

\subsection*{3. The Battery--Discharge--Model}

This section introduces the Battery--Discharge--Model (BDM), a stochastic
infinite state machine or Markov chain, which will serve as a
container to memorize the behaviour of $\deg$ in the mSCFA for 
{\it   all} inputs  $a\in\fqw$ simultaneously.

Since the linear complexity grows approximately like 
$L_a(n) = \deg(n,M)\approx \left\lceil
n\cdot \frac{M}{M+1}\right\rceil$  (exactly, if we had always
$\delta(n,m)\neq 0$),  and the  auxiliary degrees $w_m$ of the mSCFA
grow like  
$w_m\approx\left\lfloor\frac{n}{M+1}\right\rfloor$, we 
extract the {\it deviation} from this average behaviour as follows:
\bde
The {\it linear complexity deviation} or degree deviation is
\begin{align}
d := d_a(n) := \deg - \left\lceil n\cdot\frac{M}{M+1}\right\rceil\in\zz,
\end{align}
which we call the ``{\it drain}'' value, 
and the deviation of the auxiliary degrees is  
\begin{align}
 b_m :=  \left\lfloor n\cdot\frac{1}{M+1}
  \right\rfloor-w_m\in\zz,\ \ \ 1\leq m\leq M,
\end{align}
which we call the  ``{\it battery charges}''. 
\ede

The BDM will assemble all necessary information about the development
of $d_a(n)$ with $n\to \infty$ and  $a$ running through all of $\fqw$.

We establish the behaviour of $d$ and $b_m$ in two steps. First we
treat the change of $d,b_m$ when increasing $n$ to $n+1$ in the mSCFA
(keeping $\deg, w_m$ fixed for the moment):

\bq
\deg- \left\lceil (n+1)\cdot\frac{M}{M+1}\right\rceil=\left\{
\ba{ll}
\deg - \left\lceil n\cdot\frac{M}{M+1}\right\rceil-1,&n \not\equiv M
\operatorname{mod} M+1,\\ 
\deg - \left\lceil n\cdot\frac{M}{M+1}\right\rceil,  &n \equiv M
\operatorname{mod} M+1,\\ 
\ea
\right.
\eq
and
\bq
\left\lfloor (n+1)\cdot\frac{1}{M+1}\right\rfloor-w_m=\left\{
\begin{array}{ll}
\left\lfloor n\cdot\frac{1}{M+1}  \right\rfloor-w_m,&n \not\equiv M
\operatorname{mod} M+1,\\ 
\left\lfloor n\cdot\frac{1}{M+1}  \right\rfloor-w_m+1,&n \equiv M
\operatorname{mod} M+1.\\ 
\end{array}
\right.
\eq

Hence, by $(3)$ we have to decrease $d$ in all steps (we call this an
action $d_-$), except when 
$n\equiv M \to$
$n\equiv 0 \operatorname{mod} (M+1)$, 
and only here we increase all $M$ battery values $b_m$, by~$(4)$
(action $b_+$).

With $d(M,0)=b_m(M,0):=0,\forall m$, we obtain the invariant
\bq
d(n,m) + \left(\sum_{k=1}^M b_k(n,m)\right)+ n\operatorname{mod} (M+1)=0,\ \
  \forall n\in\nn_0, 1\leq m\leq M
\eq
for the initial timestep $(n,m)=(M,0)$. Also, by (3) and $(4)$, the
actions $d_-$ (increase $n$, decrease $d$) and $b_+$ (decrease $n \mod
(M+1) $ by $M$, increase $M$ batteries by $1$ each)
do not change the invariant. 

Now, for $n$ fixed, the $M$ steps of the inner loop of the mSCFA
change $w_m$ and $\deg$  only in the case of $\delta(n,m)\neq 0$ and 
$n-\deg-w_m>0$ that is 
\[n-\deg-w_m>0\stackrel{(1;2)}{\Longleftrightarrow} 
n-\left(d+ \left\lceil n\cdot\frac{M}{M+1}\right\rceil\right) 
-\left( \left\lfloor n\cdot\frac{1}{M+1} \right\rfloor-b_m\right) >0\] 
$\Longleftrightarrow b_m >d.$ 
In the case $\delta\neq 0$ and $b_m>d$, the new values are (see mSCFA)
\bq
\deg^+=n-w_m\hspace{2 cm}\mbox{\rm \ and \ }\hspace{2 cm}w_m^+ = n-\deg
\eq
and thus in terms of the BDM variables:
\bqa
d^+ \stackrel{(1;6)}{=} 
(n-w_m) - \left\lceil\frac{n\cdot M}{M+1}\right\rceil
\stackrel{(2)}{=} 
\left\lfloor\frac{n}{M+1}\right\rfloor + b_m 
- \left\lfloor\frac{n}{M+1}\right\rfloor = b_m
\eqa
and
\bqa
b^+_m \stackrel{(2;6)}{=} 
\left\lfloor\frac{n}{M+1}\right\rfloor - (n - \deg)
\stackrel{(1)}{=} 
-\left\lceil\frac{n\cdot M}{M+1}\right\rceil+ \left(d 
+\left\lceil\frac{n\cdot M}{M+1}\right\rceil\right)
=d,
\eqa
an interchange of the values $d$ and $b_m$. We say in this case that
``battery $b_m$ {\it discharges} the excess charge  into the drain'',
and call this behaviour an action ``D'' of battery $b_m$,
corresponding to case 2b of \cite[Thm.~2]{Dai}.
A discharge does not affect the invariant $(5)$, which is thus valid
for all timesteps $(n,m)$.

The remaining cases are $b_m>d$, but $\delta=0$,  an {\it inhibition}
of $b_m$, action ``I'' (case 2a  of \cite[Thm.~2]{Dai}), 
and two actions of do nothing,  ``N$_=$'' and
``N$_<$'', distinguishing  between $b_m=d$ and $b_m<d$
(case 2c and part of case 2a). 

Since we do not actually compute the discrepancy  $\delta$ (in fact,
we do not even have a sequence $a$), we have to model the distinction between
$\delta=0 $ and $\delta\neq 0$ probabilistically.

\begin{proposition}
In any given position $(n,m), n\in\nn, 1\leq m\leq M$ of the formal
power series, 
exactly one choice for the next symbol  $a_{n,m}$
will yield a discrepancy $\delta=0$, all other $q-1$ symbols from $\ff_q$ 
result in some  $\delta\neq 0$.
\end{proposition}

\begin{proof}\ 
The current approximation $u_m^{(n,m)}(x)/v^{(n,m)}(x)$ determines
exactly  {\it one } approximating coefficient sequence  for the $m$--th
formal power series $G_m$. 
The (only) corresponding symbol belongs to $\delta=0$.
\end{proof}

In fact, for every position 
$(n,m)$, each discrepancy value $\delta \in\ff_q$
occurs exactly once for some  $a_{n,m}\in\ff_q$, 
in other words (see \cite{CV}\cite{VSETA} for $M=1$):
\\\\
{\bf Fact}
\ \ 
{\it The  mSCFA induces an isometry on $\fqw$.
}
\\\\
Hence, we can model $\delta=0$ as occurring with
probability $1/q$, and  $\delta\neq 0$ as having probability $(q-1)/q$.

To keep track of the variables $d,b_m$, we define the following state
set for the BDM:

\bde
The augmented state set is
\[\overline S := \{s=(b_1,\dots,b_M,d;T,t)\ | \ b_m\in \zz, 1\leq m\leq M; 
d\in\zz;\hspace*{4 cm}\]
\[\hspace*{4 cm}T\in\zz; 1\leq t\leq M+1; d+T+\sum_{m=1}^M b_m=0\},\]
where the last condition is the invariant $(5)$. For the BDM, we only
use the timesteps $0\leq T\leq M$, and the BDM thus has the state set  
$$S := \{s\in\overline S\ | \ 0\leq T \leq M\}$$
with  initial state $s_0:=(0,\dots,0;0,M+1)$.
\ede

To facilitate notation, we also define 
$S(T_0,t_0,d_0)=\{s\in S\ | \ T(s)=T_0, t(s)=t_0, d(s)=d_0\}$, and
similarly $S(T_0,t_0), S(T_0), \overline S(T_0,t_0,d_0)$.

A state stores the values of the batteries and the drain in 
$b_1,\dots, b_M,d$, the value $T$ corresponds
to the time modulo $M+1$ that is $T\equiv n \operatorname{mod} M+1$, and the
``ministeps'' $t=1,\dots,M$ correspond to the update of battery
$b_m$ between $t=m$ and $t=m+1$, while $t=M+1\to 1$ corresponds to the updates 
$d_-,b_+$.

The allowed transitions $\alpha$ (action) from a state
$s=(b_1,\dots,b_M,d;T,t)$ are 
$\alpha=d_-$ or $b_+$ for $t=M+1$, and otherwise depend on the relative size
of $b_t$ and $d$ ($\alpha\in\{D,I,N_=,N_<\}$). 
We have  $s\stackrel{\alpha}{\longrightarrow} s^+$ 
with the following   actions, conditions,
nextstates $s^+$, and probabilities:

\noindent
$
\ba{lllc}
\alpha&\mbox{condition}&s^+&\mbox{prob.}\\
D  &b_t>d,t\leq M&(b_1,\dots,b_{t-1},d,b_{t+1},\dots,b_M,b_t;T,t+1)
&{q-1}/{q}\\   
I  &b_t>d,t\leq M&(b_1,\dots,b_M,d;T,t+1)    &{1}/{q}\\
N_=&b_t=d,t\leq M&(b_1,\dots,b_M,d;T,t+1)   &1\\
N_<&b_t<d,t\leq M&(b_1,\dots,b_M,d;T,t+1)   &1\\
d_-&T<M,t=M+1&(b_1,\dots,b_M,d-1;T+1,1) &1\\
b_+&T=M,t=M+1&(b_1+1,b_2+1,\dots,b_M+1,d;0,1)&1\\
\ea
$

Whenever $b_t > d$, both D and I may occur, leading to two
feasible transitions from  a given state $s$, whose 
probabilities sum up to 1. 

Recall that from $(T,M+1)$ to $(T+1,1)$, the drain $d$ is decremented
according to 
$(3)$  for $T<M$, action ``d$_-$'', and from $(M,M+1)$ to $(0,1)$, the
batteries $b_m$ are incremented  according to $(4)$, action ``b$_+$''. 

\bde
The {\it state transition matrix} $\cal T$ of the BDM is  an infinite
stochastic matrix indexed by $s,s'\in S$, and where
\[{\cal T}(s,s') =
\left\{
\ba{ll}
1,       
&s\stackrel{N_=}{\longrightarrow}s', 
 s\stackrel{N_<}{\longrightarrow}s', 
 s\stackrel{d_-}{\longrightarrow}s', 
\mbox{\ or\ }
 s\stackrel{b_+}{\longrightarrow}s',\\ 
(q-1)/q, &s\stackrel{D}{\longrightarrow}s',\\
1/q,     &s\stackrel{I}{\longrightarrow}s',\\
0,&\mbox{\rm\ otherwise}.\\
\ea
\right.\]

Every row either includes an ``I'' and a ``D'', or else one of
``N$_=$'',  ``N$_<$'', ``d$_-$'', or ``b$_+$''. Reading the feasible
transitions backwards, one obtains that a state with $b_t<d$ 
(at $(T,t+1)$) is reached either by a discharge, or by a ``N$_<$'',
hence the corresponding column of $s'$ sums up to $\frac{q-1}{q} +1$.
Likewise, if $s'$ has $b_t>d$, this may only be the result of an
inhibition, hence column sum $1/q$. The cases ``N$_=$'', ``d$_-$'', and
``b$_+$'' all are by themselves the only nonzero entry within a
column, which has thus  sum 1.  
\ede

In terms of $d,b_m$, we have the following equivalent
probabilistic formulation of the mSCFA (timestep $t=M+1$ comes after
the {\tt FOR} $m\equiv t$ loop):

\begin{algorithm}
\noindent {\tt BatteryDischargeModel}\\
$d:=0; b_m:=0,1\leq m\leq M$\\
$d := d -1$\ // $d_-$\\ 
{\tt FOR} $n := 1,2,\dots$\\
\hspace*{6 mm}    {\tt FOR} $m:=1,\dots,M$\\
\hspace*{12 mm}        {\tt IF} $b_m > d$:\\
\hspace*{18 mm}            {\tt WITH} prob.~$(q-1)/q$:\\ 
\hspace*{24 mm}  {\tt swap}$(b_m,d)$  // action $D$\\
\hspace*{18 mm}            {\tt WITH} prob.~$1/q$:\\
\hspace*{24 mm}     \{\}  // action $I$\\
\hspace*{12 mm}        {\tt ELSE}\\
\hspace*{18 mm}      \{\}  // action $N_=,N_<$\\
\hspace*{12 mm}         {\tt ENDIF}\\
\hspace*{6 mm}     {\tt ENDFOR}\\
\hspace*{6 mm}  {\tt IF} $n\not\equiv M \operatorname{mod} M+1:$\\
\hspace*{12 mm} $d := d -1$\ \ // $ d_-$\\ 
\hspace*{6 mm}  {\tt ELSE}\\     
\hspace*{12 mm} $b_m := b_m + 1, 1\leq m \leq M$\ // $b_+$\\ 
\hspace*{6 mm}          {\tt ENDIF}\\
{\tt ENDFOR}
\end{algorithm}

\subsection*{4. Classes of BDM States}

The Markov chain BDM will turn out to be strongly concentrated on few states. 
We define a family of measures $\mu_\tau$ on $S$, indexed by 
$\tau\in  \nn_0$. We start for $\tau=0$ with all mass concentrated on the
initial state  $s_0$:
\[
\mu_0(s) = \left\{
\ba{ll}
1,&s=s_0=(0,\dots,0;0,M+1)\\
0,&s\neq s_0\\
\ea
\right.\]

For successive timesteps $\tau$, we then put $\mu_{\tau+1}(s') =
\mu_\tau(s)$, if  
$s\stackrel{N_=}{\longrightarrow} s'$,
$s\stackrel{d_-}{\longrightarrow} s'$, or
$s\stackrel{b_+}{\longrightarrow} s'$. 
Also, $\mu_{\tau+1}(s') = \frac{1}{q}\mu_\tau(s)$, if 
$s\stackrel{I}{\longrightarrow} s'$.
Finally $\mu_{\tau+1}(s') = \frac{q-1}{q}\mu_\tau(s_1)+\mu_\tau(s_2)$, if 
$s_1\stackrel{D}{\longrightarrow} s'$ and 
$s_2\stackrel{N_<}{\longrightarrow} s'$. 

Put in other words, $\ds \left(\mu_{\tau+1}(s)\right)_{s\in S} 
= {\cal T} \cdot \left(\mu_{\tau}(s)\right)_{s\in S}.$

Be aware that from $\mu_\tau$ to $\mu_{\tau+1}$, we only deal with
{\it one} input symbol (or $d_-,b_+$), 
hence the distribution after reading all $M$  inputs of column $n$ is in
fact $\mu_{(M+1)\cdot n}(s)$.

\bde
We will use repeatedly the ``timesteps''
$(T,t)\in\{0,\dots,M\}\times\{1,\dots,M+1\}$ of the BDM, comparing
them with linear time $\tau\in\nn_0$. We define:
\[(T,t)\equiv \tau :\Longleftrightarrow (T-1)\cdot(M+1) + t
\equiv \tau \mod (M+1)^2\]
\ede

When dealing with the $m$--th symbol in column $n$, the $\tau$--th
input symbol, we are in a state
with $T(s)\equiv n \mod (M+1)$, $t=m$, and $(T,t)\equiv \tau$.

\bpro
For every $\tau\in\nn_0$,
$$
\sum_{s\in S(T_0,t_0)} \mu_\tau(s)=
\left\{
\ba{ll}
1,&(T_0,t_0)\equiv \tau \\  
0,&(T_0,t_0)\not\equiv \tau\\
\ea\right.
$$
\epro

\bp
By induction on $(T,t)$: Initially ($\tau=0, T=0, t=M+1$), all mass is on
$s_0$. Also, every transition goes from states with  $(T,t)\equiv \tau$
to states with $(T',t')\equiv \tau+1$, carrying over the mass to the
new $S(T',t')$.
\ep

\bde
Denote the number of sequence prefixes in $\left(\ff_q^M\right)^n$
with linear complexity deviation $d\in \zz$ as  $N(n,d;q)$.
\ede

Since the BDM has been derived from the behaviour of the mSCFA,
we obtain

\bthm
Assume that exactly $N$ of the $q^{M\cdot n}$ sequence prefixes of
length $n$ lead to a certain configuration $(\deg, w_1,\dots ,w_M)$ of
the mSCFA, and let $b_1,\dots,b_M,d,T$ be derived from $n,\deg, w_m$
according to $(1),(2)$, then 
\[\mu_{(M+1)\cdot n}(b_1,\dots,b_M,d;T,M+1) = \frac{N}{q^{M\cdot n}}\]
with $(T,M+1)\equiv  (M+1)\cdot n$ that is
$T\equiv n\mod (M+1)$.
\ethm

\bp
The theorem is true for $n=0$, $(T,t)=(0,M+1)$ 
with $N=1$, $\deg=d=b_m=w_m=0,\forall
m$, starting with the (only) prefix $\varepsilon$, the empty string.

From then on, by the construction of the BDM, for $t(s)\leq M$  a transition 
$s\to s^*$ takes place with probability $\frac{b}{q}$, with $b$ from
$\{1,q-1,q\}$, if and only if the mSCFA goes to the state
corresponding to $s^*$ for $b$ out of the $q$ possible next symbols
$a_{n,m}\in\ff_q$,
or, for $t(s)=M+1$, $t(s^*)=1$, corresponding to  actions $d_-, b_+$, 
with probability one, while the mSCFA increases $n$.    
\ep

From this theorem now follows as a corollary the description of
$N(n,d;q)$ by the mass distibution on the BDM states
(more on finite $n$ in Section 9):

\bthm For $n\in\nn_0$ with $(T_0,M+1)\equiv (M+1)\cdot n$, 
for $d\in \zz$,
$$N(n,d;q) = q^{M\cdot n} \times \sum_{s\in S(T_0,M+1,d)}
\mu_{(M+1)\cdot n}(s)$$  
\ethm

\bde
For a given state $s\in S$, we define its asymptotic measure as  
\[\mu_\infty(s) := \limsup_{n\to\infty}\mu_n(s)
= \lim_{
\scriptsize
\ba{c}
n\to\infty\\
n\equiv (T(s),t(s))
\ea
}\mu_n(s).\] 
We have $\sum_{s\in S} \mu_\infty(s) = (M+1)^2$, since each $S(T,t)$ 
sums up to 1.
\ede

We will see that all states satisfy
$\mu_{\infty}(s) = \mu_{\infty}(s_0) \cdot q^{-K(s)}$ 
for some $K(s)\in\nn_0$.
We call this value $K(s)$ the {\it class} of state $s$ and define it
algorithmically, generalizing to $s\in \overline{S}$:

\bde
The {\it class} of a state $s=(b_1,\dots,b_M,d;T,t)\in\overline S$ is 
$$K(s) = -\pi_s + M\cdot T +2\cdot \sum_{m=1}^{M+1} {\tilde b}_m\cdot
(M+1-m),$$
where $\pi_s$ is minimum  number  of  transpositions between neighbours
necessary to sort $(b_1,\dots,b_{t-1},d,b_t,\dots,b_M)$ into
decreasing order as $({\tilde b}_1,\dots,{\tilde b}_{M+1})$,
${\tilde b}_i\geq {\tilde b}_{i+1}, 1\leq i\leq M)$.
Observe that the place of $d$ in the initial sequence depends on $t$.
\ede

\begin{example}
The state $s=(-5,4,-4,2;1,2)$ with $M=3$, $d=2$, $T=1$ and $t=2$
requires the sorting of $(-5,2,4,-4)$ into $(4,2,-4,-5)$, using
$\pi_s=4$ transpositions, and thus 
$K(s) = -4+3\cdot 1+2(4\cdot 3+2\cdot
2+(-4)\cdot 1 + (-5)\cdot 0) = 23$.
\end{example}   

This {\it static} way of determining $K(s)$ is compatible with the
following {\it dynamic} consideration of transitions. First we need a
technical lemma:

\blem
For all $s=(b_1,\dots,b_M,d;T,t)\in \overline S$, we have
$$K(s) = K(b_1,\dots,b_M,d;T,t) = K(b_1+1,\dots,b_m+1,d+1;T-M-1,t).$$
\elem

\bp
Let $s=(b_1,\dots,b_M,d;T,t)$ and 
$s'=(b'_1,\dots,b'_m,d';T',t):=(b_1+1,\dots,b_m+1,d+1;T-M-1,t)$.
Then $\pi_s = \pi_s'$, since the relative order within $s$ and $s'$
are the same, also $\tilde b_i'=\tilde b_i+1,  1\leq i\leq M+1$. 
Using 
\[2\cdot \sum_{m=1}^{M+1} 1 \cdot(M+1-m)=2
\genfrac(){0cm}{}{M+1}2 =
M\cdot(M+1),\] 
we thus have 
\bqa
K(s) 
&=& -\pi_s + M\cdot T +2\cdot \sum_{m=1}^{M+1} {\tilde b}_m\cdot(M+1-m)\\
&=& -\pi_s' + MT-M(M+1)
+2\cdot \sum_{m=1}^{M+1} ({\tilde b}_m+1)\cdot(M+1-m)\\
&=& -\pi_s' + M\cdot T' +2\cdot \sum_{m=1}^{M+1}{\tilde
  b}'_m\cdot(M+1-m)=K(s'). \qedhere
\eqa
\ep

We now obtain the change in class by counting actions $I$ and $N_<$:

\bthm \hspace*{2 mm}\\
\indent$(i)$ For every feasible transition $s\stackrel {\alpha}{\to}
s'$ between 
states $s,s'\in S$ with $\alpha\in\{D,I,N_=,N_<,d_-,b_+\}$, we have  
\[K(s') = K(s) + \left\{
\ba{rl}
1,&\alpha=I\\
-1,&\alpha=N_<\\
0,&\alpha \in\{D,N_=,d_-,b_+\}\\
\ea
\right. 
\]

$(ii)$ Let $s_0\stackrel{\alpha_1\dots\alpha_k}{\longrightarrow}s$ be some
path from the initial state $s_0$ to $s$.\\ 
Let $\#I=\#\{1\leq i \leq k\ | \alpha_i=I\}$ and
$\#{N_<}=\#\{1\leq i \leq k\ | \alpha_i=N_<\}$. Then
\[ K(s') - K(s) = \#I-\#{N_<}.\]
\ethm

\bp
$(i)$ We first deal with transitions from $D,I,N_=,N_<$ that is $t\neq
M+1$. Since the values of the multiset $\{b_1,\dots,b_M,d\}$ only get
swapped (in the case of a discharge D), the sum 
$2\sum_{m=1}^{M+1}\tilde b_m\cdot(M+1-m)$ 
as well as the term $M\cdot T$ stay the same. 

It suffices thus to compare $\pi_s$ with $\pi_{s'}$.
Let $(b_1,\dots,b_{t-1},d,b_t,\dots,b_M)$ be the sequence to be
ordered for $s$ and similarly for $s'$, after the discharge, 
$(b'_1,\dots,b'_{t},d',b'_{t+1},\dots,b'_M)$.
Since $b_m=b'_m$ for $m\neq t$, they need the same number of transpositions, 
and we may in fact restrict our comparison
to the sorting of $(d,b_t)$ for $s$ and $(b'_t,d)$ for $s'$.

Case $\alpha=D$: We had $b_t>d$ (to be able to apply $D$), $d' := b_t,
b'_t := d$ and thus $d'>b_t'$. Both before and after the discharge,
one transposition is necessary and thus $\pi_s=\pi_{s'}$, $K(s) =
K(s')$.

Case $\alpha=I$: Again $b_t>d$,  $b_t' := b_t, d' := d$. We sort $(d,b_t)$
with one transposition, $(b_t',d')$ is already sorted. 
Hence  $\pi_{s'}=\pi_s-1$, $K(s') = K(s)+1$.

Case $\alpha=N_=$: Now, $b_t=d=b_t'=d'$ and so  
$\pi_{s'}=\pi_s$, $K(s') = K(s)$.

Case $\alpha=N_<$: Here $b_t < d$ and $b_t'<d'$. $(d,b_t)$ is already in
order, $(b_t',d')$ requires a transposition and so $\pi_{s'}=\pi_s+1$, 
$K(s') = K(s)-1$.

Finally, for $d_-$ we have $s'=(b_1,\dots,b_M,d-1;T+1,1)$ from
$t=M+1$ for $s$. We have to sort $(b_1,\dots,b_M,d)$ for $s$ (with $t=M+1$) and
$(d-1,b_1,\dots,b_M)$ for $s'$ (with $t=1$).
Sorting first the part $b_1,\dots,b_M$ with equal effort in both $s$
and $s'$,  $\overline\pi_s =
\overline\pi_{s'}$, we then introduce $d$ from the right, respectively
$d-1$ from the left to the {\it same} place:
$(\tilde b_1,\dots,\tilde b_{k-1},d\mbox{\rm \ or \ } d-1,\tilde
b_{k+1},\dots,\tilde b_M)$, where $\tilde b_{k-1}\geq d>d-1$ and
$\tilde b_{k+1}\leq d-1<d$.
The total number of transpositions is then $\pi_s = \overline
\pi_s+(M+1-k)$ and $\pi_{s'} = \overline\pi_{s'}+(k-1)$.

The class now is
\[K(s) = -\pi(s)+M\cdot T+2{\sum_{m=1,m\neq k}^{M+1}}
\tilde b_m(M+1-m)
+2d(M+1-k) \]
\[= -\pi(s)+(M-(k-1))-(k-1)+M\cdot (T+1)\]
\[+2{\sum_{m=1,m\neq k}^{M+1}} \tilde b_m(M+1-m)
+2(d-1)(M-(k-1))=K(s').\]

The case $\alpha = b_+$ is equivalent to  $d_-$
followd by incrementing all the $b_m$ and $d$, hence 
follows from the case $\alpha=d_-$ and Lemma~7.

$(ii)$ This follows by applying $(i)$ to the $k$ transitions leading
to $s$, starting in $s_0$ with $K(s_0)=0$.
\ep

We will now show that the limit mass distribution $\mu_\infty$ follows
in fact  (up to a constant) from the state classes as
$\mu_\infty(s)=C_0\cdot q^{-K(s)}$. 
First, we state a theorem by Rosenblatt (an infinite matrix version of
Perron--Frobenius):

\bthm {\rm (Rosenblatt, \cite{Rosen})}
Let {\cal T} be a Markov chain, finite or infinite. 
``If the chain is irreducible and
nonperiodic,  there is an invariant instantaneous distribution if and
only if the states  are persistent, in which case the distribution is
unique and given by  $\{u_k\}$'' $\cite[p.~56]{Rosen}$, where 
$u_j = \lim_{n\to\infty}p_{j,j}^{(n)},$ and $p_{j,j}^{(n)}$ is the
probability to return to state $j$ after $n$ steps.
\ethm

\bp
See \cite[p.~56]{Rosen}.
\ep

Here $\cal T$ certainly {\it is} periodic, with period $(M+1)^2$.
The $(M+1)^2$--th  power of $\cal T$ has the property that transitions
occur  only
within the sets $S(T,t)$, so it can be ordered into  a block diagonal matrix.
We use only the block with $(T,t)=(0,M+1)$, including $s_0$, as
$\widehat {\cal T} := {\cal T}^{(M+1)^2}|_{s\in S(0,M+1)}$. 

${\cal T}$ and thus $\widehat {\cal T}$ is irreducible, since we get
from $s_0$ to every state and back by the following theorem:

\bthm
$(i)$ For every state $s\in S$, there is exactly one 
sequence of transitions $\underline\alpha=\alpha_1\cdots\alpha_k$ with 
$s_0\stackrel{\underline\alpha(s)}{\longrightarrow}s$ and
$\underline\alpha(s)\in\left\{D,I,N_=,d_-,b_+\right\}^*$  
$($avoiding actions of the type $N_<)$, which touches the state $s_0$ 
only initially. 

\noindent
$(ii)$ Also, there is exactly one path from $s$ to $s_0$ avoiding actions of
type $I$,  which touches the state $s_0$  only finally. 
\ethm

\bp
$(i)$
{\it Unicity}:
There is at most one such transition: When going backwards from $s$ to
$s_0$ , running through the batteries in reverse order $M,M-1,\dots,1$ 
for each transition, we have:

For $b_m < d$ this results either from a discharge $D$, or a do
nothing $N_<$. Since $N_<$ is not allowed in $a_s$, put D.

For $b_m = d$ this results from a do nothing $N_=$. 

For $b_m > d$, only  an inhibition I is possible.

{\it Existence}:
There is an infinite chain of predecessors, all of class less than or
equal to $K(s)$. Since for each $K$, there are only finitely many
states with this class, in particular, there is some state $s^*$ with
$(T,t)=(0,M+1)$, which is reached repeatedly. If this state is
$s^*=s_0$,  we are done. If not, $s^*=(b_1,\dots,b_M,d;0,M+1)$ 
has $mx:=\max\{b_1,\dots,b_M,d\}\geq 1$
and $mn:=\min\{b_1,\dots,b_M,d,\}\leq -1$ (by the invariant $(5)$, with
$T$ being zero). 

However,  a cycle
$s^*\stackrel{\{D,N_=,d_-,b_+\}^+}{\longrightarrow}s^*$  without $I$
or $N_<$ is impossible:
Either $d=mx$ at $(T,t)=(0,M+1)$, or else some battery $b_t=mx$
has to discharge ($I$ prohibited). At $(T,t)=(1,M+1)$, we have
$d=mx\geq 1$ in any case, thus at $(T,t)=(2,1)$, we get $d\geq 0$.
Now, since $mn\leq -1$ is the value of one of the batteries, say
$b_{t^*}$, at time $(2,t^*)$ we have $b_{t^*}<d$ and  thus $N_<$ is
the only possible action.
So, no return to $s^*$ avoiding $I$ and $N_<$ (having reached $K=0$,
there is no further decrement) is possible, unless
$s^*=s_0$. 
Since the only cycle to avoid passes repeatedly through $s_0$, 
$\underline \alpha(s)$ is well-defined by excluding this case.

$(ii)$ To get back, just choose $D$, whenever $b_t>d$. In this way,
the class can never increase, and thus eventually, we must hit a
cycle. But we have already seen that the only cycle avoiding both $I$
and $N_<$ passes through the states with class 0, including $s_0$.
\ep

\bthm
For any two states $s,s'\in S$, 
\[\frac{\mu_\infty(s)}{\mu_\infty(s')} = q^{K(s')-K(s)}.\]
\ethm

\bp
Let a mass distribution  $\mu(s) := q^{-K(s)}$ be given. 
We show that $\mu$ is invariant under the transition matrix of the
BDM, {\it i.e.} $(\mu(s))_{s\in S}$ is an eigenvector of eigenvalue one,  
and unique with this property up to a constant factor.
We consider all states leading to a fixed state $s$.
We have three cases:

1. $b_t<d$ after the action, coming from 
$s_1\stackrel{D}{\longrightarrow} s$ or
$s_2\stackrel{N_<}{\longrightarrow} s$, and thus
$\mu_\infty(s) =  \frac{q-1}{q}\mu_\infty(s_1) + \mu_\infty(s_2)$.
Since $N_<$ decrements the class (but $D$ not), we have 
$ \frac{q-1}{q}p_o\cdot q^{-K(s_1)}+p_o\cdot q^{-K(s_2)}
= \frac{q-1}{q}p_o\cdot q^{-K(s)}+p_o\cdot q^{-(K(s)+1)}
= \left(\frac{q-1}{q}+\frac{1}{q}\right)\cdot p_o\cdot q^{-K(s)}
= p_0\cdot q^{-K(s)}$. 

2. $b_t=d$ after the action, which must be a  do nothing, $\alpha=N_=$,
and thus $K(s)=K(s')$, $\mu_\infty(s) = \mu_\infty(s')$.

3. $b_t>d$ afterwards (and before), from an inhibition, $\alpha=I$ which
   increments the class, hence  $q^{-K(s)}\cdot\frac{1}{q}=q^{-(K(s)+1)}$

This shows {\it consistency} of $\mu(s) =c\cdot  q^{-K(s)}$ with the
behaviour of the BDM, or stated otherwise:
$(\mu(s))_{s\in S} = (q^{-K(s)})_{s\in S}$ is an eigenvector 
of the infinite state transition matrix of the BDM. Furthermore its
eigenvalue 1 is the {\it largest} eigenvalue of ${\cal T}$, since
${\cal T}$ is stochastic.

Now, $\widehat {\cal T}$ inherits the eigenvector $\mu$, restricted to states
from $S({0,M+1})$, with eigenvalue $1^{(M+1)^2}=1$. This matrix is
aperiodic and irreducible by Theorem 10, and by  Theorem 9
(Rosenblatt), $\mu$ is already the {\it only} 
such eigenvector up to a constant factor,  and it remains to normalize it.

Returning from $\widehat {\cal T}$ to ${\cal T}$, we obtain the
statement, since $\mu(s)=q^{-K(s)}$ for all $s\in S(0,M+1)$ forces all
other states in $S$ also into this eigenvector.
\ep

\subsection*{5. Antisymmetry}

In this section, we consider only the configurations with $t=M+1$, at
the end of a complete column from the input $a$.

\bpro
For  $M\in\nn$, $T\in\zz$,  
$d\in \zz$, $k\in\nn_0$,  and $2\leq q\in \nn$, let
$A=\{s\in \overline S({T,M+1,d})\ | \ K(s) = k\}$ and\\ 
$\overline A = \{s\in \overline S({M-T,M+1,-d})\ | \ K(s) = k\}$.
Then $|A|=|\overline A|$. 
\epro

{\it Proof}.
We show that states $s :=(b_1,\dots,b_M,-T-X;T,M+1)$ and\\ 
$\overline s :=(-b_M-1,\dots,-b_1-1,T+X;M-T,M+1)$,
where $X:= \sum_{m=1}^M b_m$, satisfy:\\
$(i)$ $T(s)=T$ and $T(\overline s) = M-T$,
$(ii)$ $d(s) = -d(\overline s)$, and
$(iii)$ $K(s) = K(\overline s)$. \\
Then $s\in A\Longleftrightarrow \overline s \in
\overline A$, and we have a  bijection
between $A$ and $\overline A$, hence $|A|=|\overline A|$.
 
$(i)$ and $(ii)$ are obvious by inspection. To show $(iii)$,
we first sort $b_1,\dots, b_M$ of $s$ into decreasing order as
$\tilde b_1\geq \tilde b_2\geq\dots\geq \tilde b_M$ by $\pi_s'$
permutations of neighbours. 
Then  $-b_M-1,\dots, -b_1-1$ of $\overline s$ can be sorted into
decreasing order 
by $\pi_s'$ permutations at the same places into
$-\tilde b_M-1,\dots,-\tilde b_1-1$.

We now introduce $d=-T-X$ and $-d=T+X$, resp., into the ordered $\tilde b$'s as
$\tilde b_1\geq\dots \tilde b_{k}\geq -T-X> \tilde b_{k+1}\geq \dots
\tilde b_M$ and  
$-\tilde b_M-1\geq\dots -\tilde b_{k+1}-1\geq T+X> -\tilde b_{k}-1\geq
\dots \tilde b_1$ (observe 
the $>$ inequality in {\it both} cases to the right of $\pm(T+X)$).
We have a total of $\pi_s = \pi_s'+M-k$ and $\pi_{\overline s} =
\pi_s'+k$, resp., permutations, thus
$\ds K(s) =-\pi_s -(M-k)+ M\cdot T
+2\sum_{i=1}^M(M+1-i)\tilde b_i-2\sum_{i=k+1}^M \tilde b_i + 2(M-k)(-T-X)$
and 
$\ds K(\overline s) = -\pi_s-k+M\cdot (M-T)
+2\sum_{i=1}^M(M+1-i)(-\tilde b_{M+1-i}-1)-2\sum_{i=1}^k (-\tilde
b_i-1) + 2k(T+X),$ 
where the first sum treats the $\tilde b_i$'s in their place {\it before}
introducing $\pm(T+X)$, the second sum adjusts the $\tilde b_i$'s,
which are shifted while introducing $d$,  by 2,
and the last term belongs to the drain $\pm(T+X)$. 
The difference is then 
\bqa
&&K(\overline s) - K(s) 
= M^2-2MT+M -2k +2\sum_{i=1}^M(-i-M-1+i)\tilde b_{i}-\\
&&-2\sum_{i=1}^M 1 +2\sum_{i=1}^M \tilde b_i
+2\sum_{i=1}^k 1 + 2(k+M-k)(X+T)\\
&=& M(M+1)-2MT -2k - 2(M+1)\sum_{i=1}^M\tilde b_{i}-M(M+1)+\\
&&+2\sum_{i=1}^M \tilde b_i+2k  +2MX+2MT = - 2(M+1)X+2X  + 2MX=0,
\eqa
and we obtain $(iii)$.
\hfill$\Box$

\bthm $($Antisymmetry$)$

For all $M\in \nn$,  $T\in \zz$, and $d\in \zz$,
\[\sum_{s\in\overline S({T,M+1,d})} q^{-K(s)}
=\sum_{s'\in\overline S({M-T,M+1,-d})} q^{-K(s')}.\]
\ethm

\bp
As in the proof of the  preceeding proposition, we can match the states in
the first 
sum with those in the second one. From property $(iii)$ in 12, we conclude
that the classes, and thus the sum terms,  are the same in each case.
\ep

\bde
For all $d\in\zz$, $q=|\ff_q|$, $M\in\nn$, and  $0\leq T\leq M$, let
$$\gamma(d,T,M+1) := \sum_{s\in S({T,M+1,d})} \mu_\infty(s)$$
be the asymptotic mass on all states with a given drain value $d$, 
at times $\equiv (T,M+1)$. 
\ede

\bthm
For all $d\in \zz$, $M\in\nn$,  and $0\leq T\leq M$, we have
$$\gamma(d,T,M+1) = \gamma(-d,M-T,M+1).$$
\ethm

\bp
This follows  immediately from Proposition 12 and
Theo\-rem 11.
\ep

\bde
For $0\leq T\leq M$ and $1\leq t \leq M+1$, let 
\[\overline d(T,t) :=\sum_{s\in S(T,t)} \mu_\infty(s)\cdot d(s).\]
Also, let 
\[\overline {\overline d}:= \frac{1}{M+1}\sum_{T=0}^M \overline d(T,M+1).\]
\ede

\bpro \hspace*{ 1 cm}\\
$(i)$ For $0\leq T\leq M$, $\overline d(T,M+1) = -\overline d(M-T,M+1)$.

\noindent
$(ii)$ For even $M$, we have $\overline d(M/2,M+1)=0$.
\epro

\bp
$(i)$ follows from Theorem 14, since $\overline
d(T,M+1)=\sum_{d\in\zz} d\cdot \gamma(d,T,M+1))$.\\
$(ii)$ follows from $(i)$ with
$\overline d(M/2,M+1)=-\overline d(M/2,M+1)\Rightarrow \overline
d(M/2,M+1)=0$.  
\ep

\bthm
For every $M\in \nn$, $\overline  {\overline d}=0$.
\ethm

\bp
Using $15(i)$ (and $15(ii)$ in case of even $M$), we have
\[\overline  {\overline d}
= \frac{1}{M+1}\sum_{T=0}^M \overline d(T,M+1)
= \frac{1}{M+1}\left(\sum_{T=0}^{\lfloor M/2\rfloor} \overline
d(T,M+1)+\overline d(M-T,M+1)\right)\]
=0.
\ep

{\it Remark.} In particular, Theorem 16  is an (aesthetical)
reason to choose  
$L_a(n) \approx\lceil n \cdot \frac{M}{M+1}\rceil$ (and {\it not}  
$L_a(n) \approx  n \cdot \frac{M}{M+1}$) as ``typical'' average behaviour,
another reason is that for $q\to\infty$ this same
$\lceil\dots\rceil$ value is the limit behaviour.

\subsection*{6. The Partition Model}

\bde
Let $P_M(K)\in\nn$, for $ M\in\nn,K\in\nn_0$, 
be the number of partitions of $K$ into
at most $M$ parts $($equivalently, into parts of size at most~$M)$.  
\ede

\bde
Let ${\cal P}(M,q)=\sum_{K=0}^\infty P_M(K) \cdot q^{-K}$.
\ede

\bpro a$)$  The following initial values and recursion formulae hold:
$P_1(K) =1, \forall K\in\nn$, 
$P_M(1) = 1, P_M(K)=0,  \forall K\leq 0, \forall  M\in\nn$, 
and\\  
$P_M(K) = P_M(K-M) + P_{M-1}(K)$.

\noindent b$)$ The generating function of $P_M(K)$ in powers of $q^{-1}$ is
$${\cal P}(M,q)=\sum_{K=0}^\infty P_M(K) \cdot q^{-K} =\prod_{m=1}^M
\frac{q^m}{q^m-1}.$$ 

\noindent c$)$ 
$\ds P_M(K) \approx\frac{K^{M-1}}{M!(M-1)!}$ for fixed $M$ and
$K\to\infty$. 
\epro

\bp
See \cite{CRC}, Sections 2.5.10, 2.5.12 and 2.5.11. 
\ep

\begin{remark}
Observe that by c),  for every $K\in\nn_0$, we have only 
{\it polynomially} many states of class $K$, each with  
{\it exponentially small} probability $q^{-K}\cdot \mu(s_0)$. 
This leads to the concentration of mass on the states with small $K$.
\end{remark}

\bde
Let $I_m(s), 1\leq m\leq M$, count the number of actions $I$ at battery
$m$ during  $\underline\alpha_s$ (see Theorem 10).
If $K(s)=0$, put $I_m(s)=0, 1\leq m\leq M$.

Let $(\tilde I_1,\dots,\tilde I_M)$ be the ordered 
$(\tilde I_i \geq \tilde I_{i+1}, 1\leq i <M)$  version of
$\{I_m\}$. 
\ede

\bcor
Let $\#I$ be the number of inhibitions during all of the transitions
in $\underline\alpha(s)$, similarly $\#N_<$. 
Then $\ds\sum_{m=1}^M I_m = \#{I} = K(s).$
\ecor

\bp
The $I_m$ sum up to $\#{I}$ by definition.
By Theorem 8, we have $K(s) - K(s_0) = \#{I} -
\#{N_<}$. With $\#{N_<}=0$ and $K(s_0)=0$, $\#{I} = K(s)$ follows. 
\ep

\bthm \hspace*{1 mm}\\
$(i)$ For $1\leq M\leq 8, 0\leq T\leq M, 1\leq t \leq M+1$, and 
$0\leq K\leq 1200-100M$, the state set
$S(T,t)$ contains exactly $P_M(K)$ states with $K(s)=K$.

$(ii)$ For $1\leq M\leq 8$
and 
$0\leq K\leq 600-50M$, 
fix a time $(T_0,t_0)$.
Then the $(\tilde I_1,\dots,\tilde I_M)$ of all the $P_M(K)$ states in 
$S(T_0,t_0)$  with $K(s)=K$ give the $P_M(K)$ different partitions of
$K$ into $M$ parts $($including those of size 0$)$.
\ethm

\bp
By numerical simulation over the mentioned ranges.
\ep

\bcon
The previous theorem holds for every $M,T,t,K$.
\econ

\newpage
\noindent
A graph for $M=2$, showing states $(b_1,b_2,d)$  with $K(s)\leq 4$ and\\
their associated partitions:
\\\\
\includegraphics{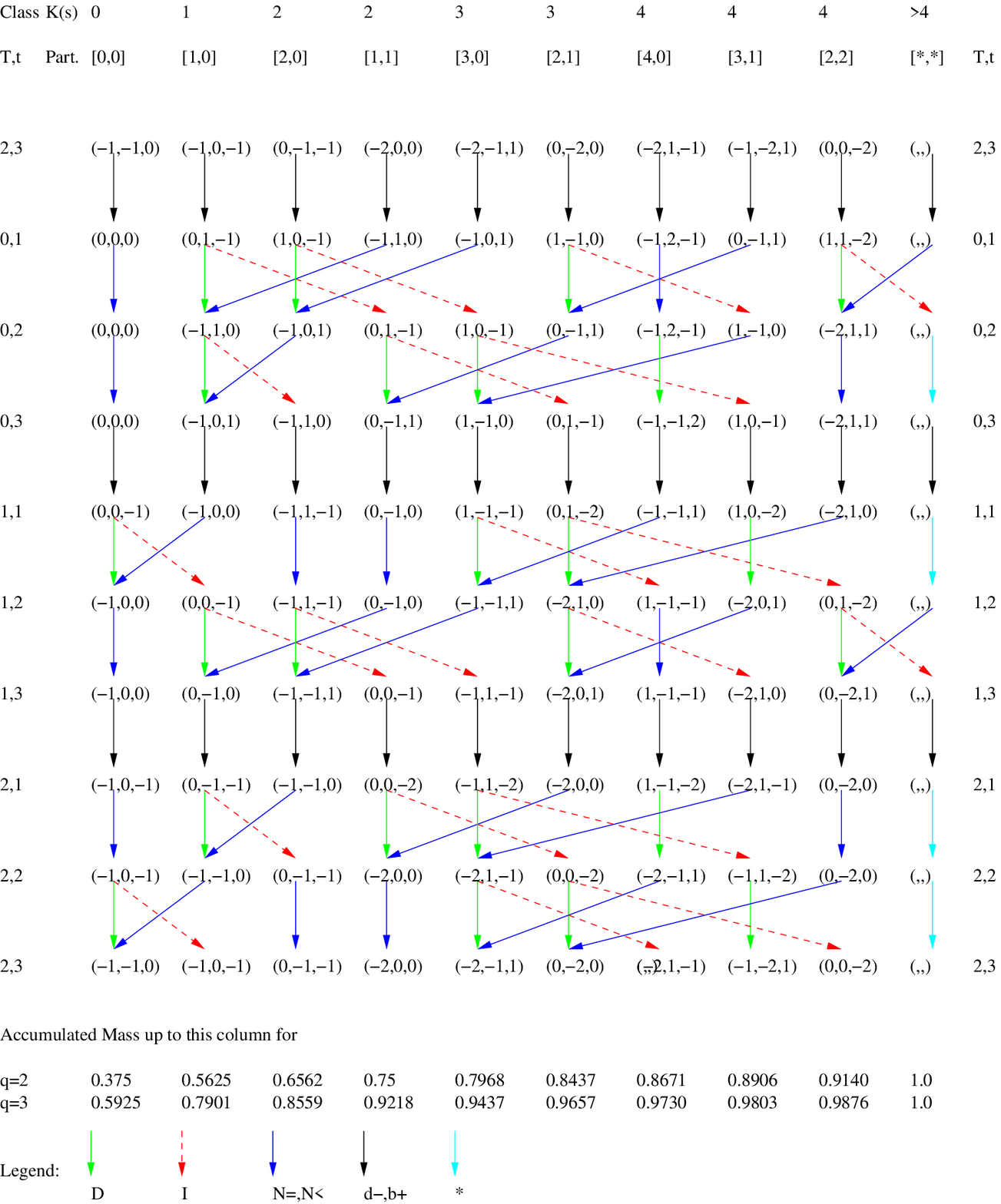}

\bcon ({\bf Theorem} for $M\leq 8$)\ \
For every state $s\in S$,
$$\mu_\infty(s) =\frac{q^{-K(s)}}{{\cal P}(M,q)}.$$
\econ

\bp
We assume the previous Theorem 19 or  Conjecture 20.
To normalize, we want to have 
$$1=\sum_{s\in S} c_0\cdot q^{-K(s)} 
= c_0 \sum_{K\in\nn_0} P_M(K)\cdot q^{-K}
=c_0{\cal P}(M,q).$$

With $c_0  := {\cal P}(M,q)^{-1}=\mu_\infty(s_0)$,
$\mu_\infty(s) := q^{-K(s)}/{\cal P}(M,q)$ is a probability
distribution (with $\sum_{s\in S(T,t)}\mu_\infty(s) = 1$ 
for all $0\leq T\leq M, 1\leq t\leq M+1$), 
which is invariant under ${\cal T}.$
\ep

\subsection*{7. Asymptotic ($n\to\infty$) Measure for the\\ 
Linear  Complexity Deviation}

\bde
Let the mass on states with drain (deviation) $d$ be 
$\gamma(d,T,t)=
\sum_{s\in S({T,t,d})} \mu_\infty(s),$
distinguished according to the timesteps $(T,t)$.
\ede

Numerical results indicate that $\gamma$ indeed depends 
only on the difference $t-T$:

\bthm
For $1\leq M\leq 8$, $0\leq T\leq M$, $1\leq t\leq M+1$, 
and any finite field $\ff_q$, let $\Delta := t-T$. Then 
for every linear complexity deviation $d\in\zz$,
\bq
\gamma(d,T,t) \dot{=} \frac{1}{{\cal P}(M,q)}
\sum_{h=1}^M 
(-1)^{h+1}
\frac{\sum_{k=0}^{h-1}q^{(M+1)\cdot k}}{q^{(M+1)(h-1)}}
\cdot
\frac{q^{-h\cdot(M+1)\cdot |d|}
\cdot
q^{\varepsilon_{\operatorname{sgn}(d)}(\Delta,h)}
}
{\prod_{k=1}^{M-h}(q^k-1)
\prod_{k=M+2}^{M+h}(q^k-1)} 
,\eq
\noindent where
\[
\ba{lll}
\varepsilon_{-}(\Delta,h) &=& h(M-1+\Delta)-\genfrac(){0cm}{}h 2\\
\varepsilon_{+}(\Delta,h) &=& h(\Delta-h)+\genfrac(){0cm}{}h 2\\
\varepsilon_{0}(\Delta,h) &=& \min\{\varepsilon_{+}(\Delta,h),
\varepsilon_{-}(\Delta,h)\}
\ea
\]
depends only on the {\rm sign} of $d$, and $\dot{=}$ means
equality with precision at least $q^{-(1200-100\cdot M)}$.
\ethm

\bp
By verifying all states with class up to $1200-100\cdot M$ in the
partition model.
The left and right side coincide up to  precision
$q^{-1200+100M}$.
\ep

\begin{remark}
This involved about $2^{39}$ or half a trillion states for $M=8$.\\
We used Victor Shoup's library NTL \cite{NTL} (Thank you!).
\end{remark}

\bcon
For every $M\in\nn$, $0\leq T\leq M$, $1\leq t\leq M+1$,
and every finite field $\ff_q$, with $\Delta$ and
$\varepsilon(\Delta,h)$ as before,   
for every $d\in\zz$, we have exactly  
\[
\gamma(d,T,t) {=} 
\sum_{h=1}^M 
\frac
{
(-1)^{h+1}
\left(\sum_{k=0}^{h-1}q^{(M+1)\cdot k}\right)
\left(\prod_{k=M-h+1}^{M}(q^k-1)\right)
\cdot
q^{\varepsilon_{\operatorname{sgn}(d)}(\Delta,h)}
}
{
q^{(M+1)(h+M/2)}
\left(\prod_{k=M+2}^{M+h}(q^k-1)\right)
q^{h\cdot(M+1)\cdot |d|}
} 
,\]
the same formula as in Theorem~22, rearranged.
\econ

{\it Remark:} The resulting values $\gamma(d,T,M+1)$ for $M=2$ and
$M=3$, and  $ \overline d(T,M+1)$ for $M=2$,  
cor\-res\-pond with Niederreiter's and Wang's results 
in  \cite[Thm.~3]{WN}, \cite[Thm.~4]{WN}, and  \cite[Thm.~11]{NW1},
resp., for $n\to\infty$, see also \cite{NW}. Observe that we use 
$d = L-\left\lceil n\cdot\frac{2}{3}\right\rceil$, not $L-n\cdot\frac{2}{3}$.

\subsection*{8. The Law of the Logarithm}

We follow  the approach by Niederreiter in \cite{NRlaw} for the
case $M=1$.

\bthm
For all $M\in\nn$, for all $0\leq T \leq M$, and $1\leq t\leq M+1$,
there exists a constant
$C(M,T,t)>0$ $($independent of $d)$ such that:
\[\frac{1}{{\cal P}(M,q)}\cdot q^{-|d|\cdot (M+1)}\leq \gamma(d,T,t) \leq C(M,T,t) \cdot
q^{-|d|\cdot (M+1)},\]
that is
\[\gamma(d,T,t) =\Theta(q^{-|d|(M+1)}).\]
\ethm

\bp
Lower bound:

We distinguish cases $d < 0$, $d > 0$, and $d=0$:

a) $d<0$

Let $b := \left\lfloor\frac{-d-T}{M}\right\rfloor$ and $a :=
-d-T-M\cdot b$ that is $b=-(a+d+T)/M$. 
Then $s^*:= (b,\dots,b,b+1,\dots, b+1,d;T,t)$ with
$b_1=\cdots=b_{M-a}=b$ and $b_{M-a+1}=\cdots=b_M=b+1$ is in $S(T,t)$.

The class of $s^*$ is $K(s^*)=-\pi_s+M\cdot T + 2\sum_{k=1}^a
(b+1)(M+1-k) +2\sum_{k=a+1}^M b\cdot (M+1-k)$, 
since after the sorting, the $a$
batteries with value $b+1$ will be the largest, while $d$ is the
smallest value.

Now, sorting starts with
$b,\dots,(d),\dots,b,b+1,\dots,(d),\dots,b+1$, 
where $d$ occupies the $t$--th place from the left.
$d$ moves to the right by $M+1-t$ moves, then all the $b$'s interchange
with all the $(b+1)$'s in $a(M-a)$ transpositions, yielding 
$\pi_{s^*}= M+1-t+a(M-a)$ and class
\begin{align*}
&K(s^*)\\
=& -(M+1-t+aM-a^2)+MT+2b(M+1)M/2\\
&+2[(M+1)M/2-(M-a+1)(M-a)/2]\\
=& -M-1+t-aM+a^2+MT-(d+T+a)(M+1)\\
&+(M+1)M-(M-a+1)(M-a)\\
=&|d|(M+1)    -M-1+t-aM+a^2+MT-MT-T-aM-a\\
&+(M+1)M-(M+1)M -a^2+2aM+a\\
=&|d|(M+1) -[(M-t)+1+T]\\
\leq& |d|(M+1)
\end{align*}
and already $s^*$ alone accounts for the lower bound.

b) $d>0$

With $b$, $a$, and $s^*$  as before, sorting now leads to 
$a+a(M-a)$ transpositions, since $d$ goes to the left. As before,
\[K(s^*) 
= -a(M+1-a) +MT+2dM 
+ 2\sum_{k=2}^{a+1} (b+1)(M+1-k) 
+ 2\sum_{k=a+2}^M b\cdot (M+1-k)\]
\[
= -a(M+1-a) +MT+2dM+2b(M+1)M/2-2bM 
+ 2[M(M-1)/2-(M-a)(M-a-1)/2]\] 
\[
= -a(M+1-a) +MT+2dM-(d+T+a)(M+1)+2(d+T+a) 
+ M(M-1)-(M-a)(M-a-1) \]
\[
= -aM-a+a^2 +MT+2dM-dM-MT-aM-d-T-a+2d+2T+2a\]
\[
+M^2-M -M^2+2aM-a^2-M+a \]
\[
= dM+d+T+a-2M
= |d|(M+1)-(M-T) -(M-a)\leq |d|(M+1)\]

c) $d=0$: Let $s^*$ be the (only) state in $S(T,t)$  with
$K(s^*)=0=|d|(M+1)$.

Upper bound:

\noindent We have
\[\gamma(d,T,t) 
= \sum_{s\in S(T,t,d)} q^{-K(s)}\cdot C_0\]
(use $C_0 := {\cal P}(M,q)^{-1}$, if you trust Conjecture 21), 
and
$$K(s) = -\pi_s + M\cdot T +2\cdot \sum_{m=1}^{M+1} {\tilde b}_m\cdot
(M+1-m).$$

a) $d < 0$:

\nopagebreak
Let $\{b_1, \dots, b_M\}$ be ordered nonincreasingly as 
$\tilde b'_1\geq \dots\geq \tilde b'_M$ 
(where $'$ indicates that $d$ does not enter into the sort).

We write the battery values as sum of their differences
$\Delta_k := \tilde b'_{M-k} -  \tilde b'_{M-k+1}\geq 0, 1\leq k\leq M-1$ 
\bq
\tilde b'_m = \tilde b'_M + \sum_{k=1}^{M-m} \Delta_k
\eq
(where for $m=M$ the empty sum is zero).

By the invariant $d+T+\sum_m b_m = 0$ we must have
\bq
&&d+T+\sum_{m=1}^M (\tilde b'_M + \sum_{k=1}^{M-m} \Delta_k) = 0\\
&\Longleftrightarrow&M\cdot \tilde b'_M= - d-T-\sum_{k=1}^{M-1} \Delta_k(M-k)\\
&\Longleftrightarrow&\tilde b'_M= -\frac{d}{M}
-\frac{T+\sum_{k=1}^{M-1} \Delta_k(M-k)}{M}
\eq

When running $\Delta_1,\dots,\Delta_{M-1}$  through all values from $\nn_0$
and setting $\tilde b'_M$ by (9--11) and then $\tilde b'_m$ by (8), we
obtain all possible values for $(\tilde b'_1,\dots,\tilde b'_M)$ (and
a lot more, since $\tilde b'_M$ is taken from $\frac{\zz}{M}\supset \zz$).

Furthermore, each $(\tilde b'_1,\dots,\tilde b'_M)$ corresponds to up
to $M!$ states (with different permutation of the values) with
$\{\tilde b'_m\}=\{b_m\}$.

$\pi_s$ can be bounded in general by $0\leq \pi_s\leq
\genfrac(){0cm}{} {M+1}2$
(the maximum being attained in the case of $b_1<b_2<\dots <b_m$).

The transition from $(\tilde b'_1,\dots,\tilde b'_M), d$ to 
$(\tilde b_1,\dots,\tilde b_{M+1})$ {\it i.e.} including $d$ in the
sort order, gives the inequality
$$2\sum_{m=1}^M \tilde b'_m(M+1-m) + 2\cdot 0\cdot d \leq 
2\sum_{m=1}^{M+1} \tilde b_m(M+1-m)$$
since at any rate a smaller $\tilde b'_m$ will be replaced by a larger 
$d$ or $\tilde b'_{m-1}$.

Putting things together, we have the upper bound
\bqa
\gamma(d,T,t) &\leq& C_0 
\sum_{\Delta_1\in \nn_0}\dots \sum_{\Delta_{M-1}\in \nn_0}
M!\cdot q^{\genfrac(){0cm}{}{M+1}2}\cdot q^{-MT}\times\\
&&\times \  q^{-2\sum_{m=1}^M
  -\frac{d}{M}(M+1-m)
  -2\sum_{m=1}^M\frac{-T-\sum_{k=1}^{M-1}\Delta_k(M-k)}{M}}\\
&=&-q^{-2\sum_{m=1}^M  -\frac{d}{M}(M+1-m)}\cdot C(M,T)\\
&=&- q^{-|d|(M+1)}\cdot C(M,T)\\
\eqa
where 
\[C(M,T) = C_0 
\sum_{\Delta_1\in \nn_0}\dots \sum_{\Delta_{M-1}\in \nn_0}
M!\times\]
\[\times q^{\genfrac(){0cm}{}{M+1}2}\cdot q^{-MT
  -2\sum_{m=1}^M\frac{-T-\sum_{k=1}^{M-1}\Delta_k(M-k)}{M}}\]
is independent of $d$.

b) In the case $d >0$ we follow the same idea, however we put $d$ as
first (largest) value of the sort order. We obtain
$$\tilde b'_m = \tilde b'_2 - \sum_{k=1}^{m-2} \Delta_k, 2\leq k\leq
M+1$$ 
with $\Delta_k\in\nn_0.$

The invariant $d+T+\sum_m b_m=0$ requires
$$d+\sum_{m=2}^{M+1} (\tilde b'_2-\sum_{k=1}^{m-2} \Delta_k) =0
\Longleftrightarrow
M\tilde b'_2 = -d-T+\sum_{k=1}^{M-1} \Delta_k(M-k)$$
$$\Longleftrightarrow \tilde b'_2=-\frac{d}{M}-\frac{T}{M}
+\frac{\sum_{k=1}^{M-1} \Delta_k(M-k)}{M}.$$

Again up to $M!$ states can be attached to one sorted tuple 
$(\tilde  b'_2,\dots,\tilde b'_{M+1})$, again $\pi_s\leq 
\genfrac(){0cm}{}{M+1}2$, 
and introducing $d$ (from the left) increases (if at all) the
  values, {\it i.e.} $\tilde b_m\geq \tilde b'_m$. We obtain
\bqa
&&-2\sum_{m=1}^{M+1} \tilde b_m(M+1-m)\\
&\leq&-2\cdot d\cdot M -2\sum_{m=2}^{M+1} \tilde b'_m(M+1-m)\\
&=&-2Md-2\sum_{m=2}^{M+1}\left(-\frac{d}{M}-\frac{T}{M}
+\frac{\sum_{k=1}^{m-2}\Delta_k(M-k)}{M}-\sum_{k=1}^{m-2}\Delta_k\right)
\left(M+1-m\right)\\
&=&-2Md+2\sum_{m=2}^{M+1}-\frac{d}{M}+C_1\\
&=&d(-2M+2\frac{(M-1)M}{2})+C_1\\
&=&-d(M+1)+C_1,\\
\eqa
where $C_1$ does not depend on $d$, and thus
$$\gamma(d,T,t) \leq C_0 
\sum_{\Delta_1\in \nn_0}\dots \sum_{\Delta_{M-1}\in \nn_0}
M!\cdot q^{\genfrac(){0cm}{}{M+1}2}\cdot q^{-MT} q^{-d(M+1)+C_1}$$
\[=q^{-|d|(M+1)}\cdot C(M,T).\]
\ep

\blem {\rm (Borel--Cantelli)}

$(i)$ Let $A_1,A_2,\dots$ be events which happen with probability
$a_1,a_2,\dots$, resp.

If now $\sum_{k\in \nn} a_k<\infty$, then with probability one
only finitely many of the events $A_k$ occur simultaneously.

$(ii)$ Let $A_1,A_2,\dots$ be {\underline{ independent}}  events which happen
with probability  $a_1,a_2,\dots$, resp.

If now $\sum_{k\in \nn} a_k=\infty$, then with probability one
infinitely many of the events $A_k$ occur simultaneously.
\elem

\bp
See Feller \cite[VIII.3]{Fel}.
\ep

\newpage
\bthm \hspace*{1 cm}\\ 
{\rm The Law of the Logarithm for Linear Complexity of
Multisequences}

For all $M\in\nn$ and for almost all $($in the sense of Haar
measure on $\fqw)$ sequences $a\in\left(\ff_q^M\right)^\infty$,
 we have
$$\limsup_{n\to\infty}\frac{d_a(n)}{\log n} =\ \  \frac{1}{(M+1)\log
  q}$$
and
$$\liminf_{n\to\infty}\frac{d_a(n)}{\log n} = -\frac{1}{(M+1)\log
  q}.$$
\ethm

\bp
We fix some $\varepsilon >0$ and apply the Borel--Cantelli Lemma $31(i)$
to the events
$$A_n: \left|\frac{d_a(n)}{\log n}\right|> \frac{1+\varepsilon}{(M+1)\log q}.$$

With $L := \left\lceil\frac{\log n}{(M+1)\log q}\right\rceil$, 
the probability for $A_k$ is 
\[a_k= 
\sum_{d=L }^\infty \gamma(d,T,t)
+
\sum_{d=-L}^{-\infty}   \gamma(d,T,t)
\leq 2\cdot C(M,T)
\sum_{d=L}^{\infty}
q^{-|d|(M+1)}\]
\[\qquad= 2C(M,T) \frac{q^{-L(M+1)}}{1-q^{-(M+1)}},\]
with accumulated probability
\[\sum_{n=1}^\infty a_n 
\leq \frac{2C(M,T)}{1-q^{-(M+1)}}
\sum_{n=1}^\infty q^{-(M+1)\cdot\frac{(1+\varepsilon)\log n}{(M+1)\log q}}
= \frac{2C(M,T)}{1-q^{-(M+1)}}
\sum_{n=1}^\infty n^{-(1+\varepsilon)}<\infty 
.\]

For the inner bounds, we need {\it independent} events:

Denote by $n_1,n_2,\dots$ the timesteps, when $d=0$. If this sequence
is finite, $d\to-\infty$, since at least one battery no longer
discharges. This event is of measure  zero, requiring {\it all}
discrepancies $\delta$ pertaining to that battery equal to zero from
some $n_0$ on.

Assume now an infinite sequence of these timesteps. 
Let $L_k := \left\lceil\frac{\log k}{(M+1)\log q}\right\rceil$ and let
$A_k$ be the event of $(M+1)\cdot (L+1)$ consecutive discrepancies, all
zero, after $n_k$.
The events $A_k$ are independent with probability $a_k = q^{-(M+1)(L+1)}$, 
since they belong to different, independent discrepancies.
Now, within $(L+1)(M+1)$ symbols, we have at least 
$(L+1)\frac{M+1}{M}$ columns and thus
$\lfloor(L+1)\frac{M+1}{M}\frac{M}{M+1}\rfloor\geq L$ actions $d_-$
(without intermediate discharges), and thus $A_k$ leads to $d\leq -L$.
 
With 
\[\sum_{n=1}^\infty a_n \geq
\sum_{n=1}^\infty  q^{-(M+1)(1+\frac{\log n}{(M+1)\log q})}
=q^{-(M+1)}\sum_{n=1}^\infty  n^{-1}=\infty\] 
and Lemma $31(ii)$, we get equality of the bounds.
\ep

\bcor
For all $M\in\nn$, for all $\varepsilon >0$, for almost all 
sequences from $\fqw$, it holds 
\[-\varepsilon<\liminf_{n\to\infty}\frac{d_a(n)}{n}
\leq\limsup_{n\to\infty}\frac{d_a(n)}{n}<\varepsilon\]
\ecor

\bp
Almost always, we have
\[\left|\frac{d_a(n)}{\log n}\right| \leq 
\frac{1}{(M+1)\log q}\Longleftrightarrow
\left|\frac{d_a(n)}{n}\right|\leq \frac{\log n}{n (M+1)\log q}\]
by the last theorem, and with 
\[\frac{1}{(M+1)\log q}\lim_{n\to\infty}\frac{\log n}{n}=0\]
the statement follows.
\ep

\bthm
With measure one,
\[\liminf_{n\to\infty}\frac{L_a(n)}{n}=\limsup_{n\to\infty}\frac{L_a(n)}{n}
=\frac{M}{M+1}.\]
\ethm

\bp
From $L_a(n) = d_a(n)+\left\lceil\frac{n\cdot M}{M+1}\right\rceil$ 
and the previous corollary, we have 
$$\frac{L_a(n)}{n}=\frac{d_a(n)}{n}+\frac{M}{M+1}+O(\frac{1}{n})$$ 
and thus 
\[\lim_{n\to \infty} \frac{L_a(n)}{n}=
\lim_{n\to \infty} \frac{d_a(n)}{n}+\frac{M}{M+1}=\frac{M}{M+1}.\qedhere\]
\ep

In other words, we obtain again the result of Niederreiter and Wang
\cite{NW,WN} that $\frac{L_a(n)}{n}\to\frac{M}{M+1}$ with probability one,
for all $M\in\nn$.

\newpage
\subsection*{9. Finite Strings}

\bde
For  $s\in S$, let the {\it generation} of state $s$ be
$$g(s)=\left\lceil\frac{\tilde I_1(s)}{M+1}\right\rceil\cdot (M+1).$$
\ede

%\newpage
\bcon 
For every state $s\in S$ and every $n\in\nn_0$
$$\mu_n(s) = \left\{
\ba{ll}
0,                          &n<g(s) \land (T,t)\not\equiv n\\ 
\mu_\infty(s)\cdot F(s),\ \ &n=g(s)\\
\mu_\infty(s),              &n>g(s) \land (T,t)\equiv n\\
\ea
\right.
$$
with
$$\ds F(s) = \prod_{m=M_1(s)}^M\frac{q^m}{q^m-1},$$ 
for 
$$M_1(s)=M + 1 - \#\left\{1\leq m\leq M\ |\ b_m = \max \{
b_1,\dots,b_M,d\}\right\}.$$ 

In the case of the empty product for $M_1=M+1$, $F(s) = 1$, 
and for $M_1(s) = 1,  F(s)={\cal P}(M,q)$.
\econ
\bcon
a$)$ For $g\in\nn_0$, let $\#(g,M)$ be the number of states that are
reachable in the $g$--th generation. Then
$$\#(g,M) = \genfrac(){0cm}{}{g+M}M .$$
b$)$ $\#\{s\in S\ |\ g(s)=g\} =  \genfrac(){0cm}{}{g+M}M - 
\genfrac(){0cm}{}g M.$
\econ

\subsection*{Conclusion}

We introduced the Battery--Discharge--Model BDM as a convenient container
for all information about linear complexity deviations in
$\fqw$. 

We obtained a closed formula for measures and averages 
for the linear complexity deviation, numerically proven for the cases
$M=1,\dots,8$, and conjectured for any $M$, which
coincides with the results known before for $M=1,2,3$, 
but gives a better account of the inner structure of these measures.
In particular, the measure is a sum of $M$ components of
the  form 
\[\Theta(q^{-|d|(M+1)h}), h=1,\dots,M.\]

\end{document}